\documentclass{ws-procs975x65}

\begin{document}

\title{CONSTRAINTS FROM COSMOGRAPHY IN VARIOUS PARAMETRIZATIONS}

\author{\uppercase{Alejandro Aviles$^{1,2}$}, \uppercase{Christine Gruber}$^{3}$\footnote{Email: chrisigruber@physik.fu-berlin.de},
\uppercase{Orlando Luongo}$^{1,4,5}$, \uppercase{Hernando Quevedo}$^{1,4}$}

\address{$^{1}$Instituto de Ciencias Nucleares, Universidad Nacional
Aut\'onoma de M\'exico, AP 70543, M\'exico, DF 04510, Mexico\\
$^{2}$Departamento de F\'isica, Instituto Nacional de Investigaciones Nucleares, AP 70543, \\M\'exico, DF 04510, Mexico\\
$^{3}$Institut f\"ur Theoretische Physik, Freie Universit\"at Berlin, Arnimallee 14, \\D-14195 Berlin, Germany\\
$^{4}$Dipartimento di Fisica and Icra, Universit\`a di Roma "La Sapienza", Piazzale Aldo Moro 5, I-00185, Roma, Italy\\
$^{5}$Dipartimento di Scienze Fisiche, Universit\`a di Napoli "Federico II", Via Cinthia, \\I-80126, Napoli, Italy}

\begin{abstract}
We use cosmography to present constraints on the kinematics of the Universe
without postulating any underlying theoretical model \emph{a priori}. To this end,
we use a Markov Chain Monte Carlo analysis to perform comparisons to the supernova Ia
union 2 compilation, combined with the Hubble Space Telescope measurements
of the Hubble constant, and the Hubble parameter datasets.
The cosmographic approach to our analysis is revisited and
extended for new notions of redshift presented as alternatives to the redshift
$z$. Furthermore, we introduce a new set of fitting parameters
describing the kinematical evolution of the Universe in terms of the
equation of state of the Universe and derivatives of the total
pressure. Our results are consistent with the $\Lambda$CDM model,
although alternative models, with nearly constant pressure and no
cosmological constant, match the results accurately as well.
\end{abstract}


\bodymatter
\bigskip

\section{Introduction}

In recent years, the wide success of the generally
accepted cosmological concordance model has been
overshadowed by some inconsistencies, one of which being the problem of 
dark energy, addressing the unexplained positive accelerated expansion of 
the Universe \cite{uno}. Various efforts, mostly in the form of modifications or extensions of
the standard model of cosmology, have been introduced to understand the physical
nature of dark energy. Many models have been developed in the literature, but
unfortunately none of them has managed to clarify the origin and nature of dark
energy satisfactorily \cite{Weinberg2008}. Most of these models are based on the
notion of a homogeneous and isotropic Universe, described by the Friedmann-Robertson-Walker
(FRW) metric, $ds^2=-c^2dt^2+a(t)^2(dr^2/(1-kr^2)+r^2\sin^{2}\theta d\phi^2+r^2d\theta^2)$.
Given the considerable amount of proposals to resolve the issue of
dark energy and the difficulties in distinguishing fairly between models and
evaluating precisely the degree of accordance between a model and the data, it is
desirable to develop an analysis which describes solely the kinematics of the
Universe without relying implicitly on a particular model \cite{CattViss2008}.
The purpose of this work is twofold. We first discuss the concept of
cosmography, a technique of data analysis able to fix bounds on the observable
Universe from a model-independent point of view; giving particular regard to
developing a viable cosmographic redshift parametrization, which reduces the
systematic errors on the fitting coefficients. In addition to that, we derive
constraints on the equation of state (EoS) parameter of the Universe directly from
data, alleviating the degeneracy problem between cosmological models.

\section{The experimental techniques of Cosmography}

In this section, we present the basic principles of cosmography and illustrate 
the way of performing the cosmographic analysis. By involving the cosmological
principle only, and correspondingly the FRW metric, it is possible to infer in
which way dark energy or alternative components are influencing the cosmological
evolution, without implicitly presuming any specific properties or nature of these
components. The idea is to expand the most relevant observables such as the Hubble
parameter or cosmological distances into power series, and introducing cosmological
parameters directly related to these observable quantities \cite{CattViss2008}.
In doing so, it is possible to appraise which models are well in accordance with data
and which ones should be discarded as a consequence of not satisfying the basic
demands of cosmography. In expanding the luminosity distance $d_L$ into a Taylor series 
in terms of the cosmological redshift $z$, we introduce two further notions of redshift, 
defined as $y_1 = \frac{z}{1+z}\mathrm{~~and~~}y_4=\arctan{z}$. These parameterizations 
are designed to reduce some disadvantages of the commonly
used and well-known notion of the redshift $z$ for the analysis, as e.g. the loss of
convergence of the power series for values of $z>1$. In particular, while $y_1$ was 
previously introduced in the literature \cite{CattViss2005}, 
we propose to use $y_4$, which has been obtained
by requiring a better convergence behavior of $d_L$. Our recipe for determining
a redshift variable consists in satisfying three considerations: 
$a)$ the luminosity distance should not behave
too steeply in the interval $z < 1$, $b)$ the luminosity distance should not exhibit sudden
flexes and $c)$ the curve should be one-to-one invertible. It turns out that the newly
introduced $y_4$ is more suitable for a cosmographic analysis than $y_1$. For $y_4$,
the parametrization of the luminosity distance is given by 
$d_L = c/H_0 \cdot \Bigl[ y_4 + y_4^2 \cdot \Bigl(1/2 - q_0/2 \Bigr)
+ y_4^3 \cdot \Bigl(1/6 -j_0/6 + q_0/6 + q_0^2/2 \Bigr) + \mathcal{O} (y_4^4) \Bigl]$. 
Furthermore, to counteract the problem of high inaccuracies which is created by cutting
the power series expansions too early, we have expanded all quantities up to sixth order.
For the numerical fits we made use of the recent data of Union 2 supernovae Ia,
of the Hubble Space Telescope (HST) measurements of the Hubble parameter, and of the $H(z)$
compilations \cite{tutti}, using a Markov Chain Monte Carlo method by modifying
the publicly available code CosmoMC \cite{codice}. In addition to our generalizations regarding different
notions of redshift, we also include a parametrization of the cosmological distance in terms
of the EoS parameter $\omega$ of the Universe and of the derivatives $P_i$ of the total pressure. 
This allows us to directly fit the EoS parameter of the Universe from data without having to
undergo disadvantageous error propagation in calculating the values from the cosmographic
series. This procedure gives clear constraints on the EoS parameter and on the pressure derivatives
in the framework of General Relativity, and thus provides a direct way to compare the predictions
of a model for the EoS to observational data \cite{2012Avil}. The parametrization of the luminosity
distance in terms of the EoS parameter set and as a function of $y_4$ is given by 
$d_L(y_4) = c/H_0 \cdot \Bigl[ y_4 + y_4^2/4 \cdot \Bigl( 1-3\omega \Bigr)
+ y_4^3 \cdot \Bigl( 5/24 - P_1/4 H_0^2 + \omega + 9 \omega^2/8\Bigr) + \mathcal{O} (y_4^4)\Bigr]$. 
The numerical results for the parameters of the cosmographic series, i.e. $H_0, q_0, j_0$ etc.,
using the newly introduced
redshift $y_4$, can be found in Table~\ref{tab:y4}. The numerical results show a
good agreement with $\Lambda$CDM, although they seem to be compatible with dark energy
possessing constant pressure and an evolving equation of state as well. The corresponding cosmological
model \cite{model} appears to be quite indistinguishable from $\Lambda$CDM.

\begin{table}
\tbl{Table of best fits and their likelihoods (1$\sigma$) for redshift $y_4$, for the three sets of
      parameters $\mathcal{A}\equiv \left\{H_0, q_0, j_0, s_0\right\}$, 
      $\mathcal{B}\equiv \left\{H_0, q_0, j_0, s_0,l_0\right\}$ and 
      $\mathcal{C}\equiv \left\{H_0, q_0, j_0, s_0,l_0,m_0\right\}$.
      Set 1 of observations is Union 2 + HST. Set 2 of observations is Union 2 + HST + $H(z)$.}
{\begin{tabular}{@{}ccccccc@{}}
\toprule
Parameter & $\mathcal{A}$, Set 1 & $\mathcal{A}$, Set 2 &  $\mathcal{B}$, Set 1 & $\mathcal{B}$, Set 2
& $\mathcal{C}$, Set 1 & $\mathcal{C}$, Set 2 \\
$\chi^2_{min}$ & 530.3 & 544.8 & 529.7 & 544.6 & 529.9 & 544.5 \\
\colrule
 $H_0$      & $74.55$ {\tiny${}_{-7.53}^{+7.54}$}      & $73.71$ {\tiny ${}_{-5.24}^{+5.29}$}
                    &  $73.95$ {\tiny ${}_{-7.22}^{+7.99}$}           &  $73.43$ {\tiny ${}_{-5.74}^{+6.05}$}
                    &  $74.12$ {\tiny ${}_{-7.78}^{+8.27}$}           &  $73.27$ {\tiny ${}_{-5.91}^{+6.86}$} \\

 $q_0$      & $-0.7492$ {\tiny${}_{-0.6228}^{+0.5899}$}      & $-0.6504$ {\tiny ${}_{-0.3303}^{+0.4275}$}
                    &  $-0.4611$ {\tiny ${}_{-0.6710}^{+0.5422}$}     &  $-0.7230$ {\tiny ${}_{-0.4585}^{+0.5851}$}
                    &  $-0.4842$ {\tiny ${}_{-0.9280}^{+2.7126}$}     &  $-0.7284$ {\tiny ${}_{-0.4838}^{+0.6062}$} \\

 $j_0$      &  $2.558$ {\tiny${}_{-8.913}^{+7.441}$}     &  $1.342$ {\tiny ${}_{-1.780}^{+1.391}$}
                    &  $-3.381$ {\tiny ${}_{-2.149}^{+10.613}$}       &  $2.017$ {\tiny ${}_{-3.022}^{+3.149}$}
                    &  $-1.940$ {\tiny ${}_{-2.148}^{+8.041}$}        &  $2.148$ {\tiny ${}_{-4.036}^{+3.414}$} \\

 $s_0$      &  $9.85$ {\tiny${}_{-26.69}^{+74.69}$}  & $3.151$ {\tiny ${}_{-1.771}^{+3.920}$}
                    &  $-37.67$ {\tiny ${}_{-60.10}^{+89.51}$}        &  $5.278$ {\tiny ${}_{-14.732}^{+13.076}$}
                    &  $-13.48$ {\tiny ${}_{-31.28}^{+71.65}$}        &  $2.179$ {\tiny ${}_{-35.919}^{+42.126}$} \\

 $l_0$      & --            & ---
            &  $N.C.$       &  $-0.13$ {\tiny ${}_{-65.87}^{+96.75}$}
            &  $N.C.$       &  $-11.60$ {\tiny ${}_{-187.96}^{+193.88}$} \\

 $m_0$      & --            & --
            & --            & --
            &  $N.C.$       &  $70.9$ {\tiny ${}_{-2254.5}^{+2497.8}$} \\
\botrule
\end{tabular}
}
\begin{tabnote}
$H_0$ is given in Km/s/Mpc. $N.C.$ means the results are not conclusive - the data do not constrain
the parameters sufficiently.
\end{tabnote}
\label{tab:y4}
\end{table}

\bibliographystyle{ws-procs975x65}

\end{document}